# True amplification of spin waves in magnonic nano-waveguides

H. Merbouche[1], B. Divinskiy[1], D. Gouéré[2], R. Lebrun[2], A. El-Kanj[2], V. Cros[2], P. Bortolotti[2], A. Anane[2], S. O. Demokritov[1], and V. E. Demidov[1]

[1]*Institute for Applied Physics, University of Muenster, Corrensstrasse 2-4, 48149 Muenster, Germany*

[2]*Unité Mixte de Physique, CNRS, Thales, Université Paris-Saclay, 91767, Palaiseau, France*

**Magnonic nano-devices exploit magnons - quanta of spin waves - to transmit and process information within a single integrated platform that has the potential to outperform traditional semiconductor-based electronics for low power applications. The main missing cornerstone of this information nanotechnology is an efficient scheme for the direct amplification of propagating spin waves. The recent discovery of spin-orbit torque provided an elegant mechanism for propagation losses compensation. While partial compensation of the spin-wave damping has allowed for spin-wave signal modulation, true amplification – the exponential increase in the spin-wave intensity during propagation – has so far remained elusive. Here we evidence the operating conditions to achieve unambiguous amplification using clocked nanoseconds-long spin-orbit torque pulses in sub-micrometer wide magnonic waveguides, where the effective magnetization has been engineered to be close to zero to suppress the detrimental magnon-magnon scattering. As a result, we achieve an exponential increase in the intensity of propagating spin waves up to 500 % at a propagation distance of several micrometers. These results pave the way towards the implementation of energy efficient, cascadable magnonic architectures for wave-based information processing and complex on-chip computation.**

Among novel promising nano-scale information technologies, the magnon-based information processing[1-3] occupies a special place due to the numerous advantages provided by magnetic excitations in solids – spin waves (SWs) and their quanta magnons. These excitations exist in the frequency interval from sub-GHz to sub-THz and possess wavelengths that can be as small as few tens of nanometers[4-6]. They can be guided and manipulated using simple submicrometer-wide strip magnetic waveguides and can be efficiently controlled by electric and magnetic fields, as well as by electric currents[7-12]. Thanks to these advantages, a large variety of novel magnonic devices and circuits for information processing[13-16] including neuromorphic and non-traditional computing systems[17-18] have been demonstrated in the recent years.

The main roadblock that is stalling magnonics from achieving large scale integration of chips is the lack of a reliable scheme to perform signal restauration that allows magnonic logic gates to be cascaded. Indeed, since magnons have a finite lifetime, the total length of a magnonic device cannot be much larger than their attenuation length, which is expressed as the product of the characteristic magnon lifetime and the magnon group velocity. The most promising physical phenomenon that can help overcome this difficulty, is the spin-orbit torque in layered systems constituted by a magnetic film interfaced with a conductive material with strong spin-orbit coupling[19-21]. In such systems, the electric current injected in the conducting layer is converted into a transverse pure spin current, which flows into the magnetic film. The spin current then exerts a torque on the local magnetization that counteracts the natural damping torque[22], compensating therefore the losses and eventually amplifies the magnetic excitations[23].

Although this approach seems straightforward, in practice it only works well in the limit of low currents, below the critical current, at which the complete damping compensation takes place. Such partial damping compensation can be easily achieved and was reported for standing and propagating spin waves[24-33]. In the case of propagating spin waves, it was shown to enable partial compensation of the spatial attenuation, which increases the spin-wave amplitude at a given distance from the source. However, the true amplification of spin waves by spin currents, which manifests itself by an exponential increase in the amplitude of spin waves with distance, has not yet been achieved.

Magnon-magnon interaction is the main reason why the true spin-torque amplification of SWs is difficult to achieve. In standard magnetic systems, a spin-torque induced damping compensation results in the amplification of all magnon modes, which leads to the excitation of large-amplitude coherent and/or incoherent magnetization auto-oscillations[28], which causes additional nonlinear scattering of the signal spin wave. This is the reason why in previous experiments the signal spin wave was not truly amplified, although the natural magnetic damping was compensated by the spin-torque caused by spin current[26,33].

In this work, we show that the fundamental limitations preventing the amplification of spin waves by spin currents can be overcome using a combination of nonlinearity control and temporal separation. Using microscopy imaging of propagating spin waves, we demonstrate that this approach allows one to achieve an exponential spatial increase of the amplitude of a spin wave as it propagates in a nano-fabricated waveguide based on a magnetic insulator with perpendicular anisotropy. We show that by tuning the angle of the static magnetic field, it is possible to minimize the detrimental nonlinear scattering of the signal spin wave from spin-current induced auto-oscillations. We also show that the complete elimination of the scattering, necessary to achieve the amplification, requires synchronizing the spin-current pulse with the spin wave. Otherwise said, the pulse of the signal spin wave must propagate its way before the spin current drives the spin system into a strongly excited nonlinear state. Our findings open new horizons for the development of efficient cascaded integrated magnonic information-processing circuits supported by direct on-chip spin-wave amplification.

Figures 1a and 1b show the schematics of the experiment. We study a 500-nm wide magnonic waveguide patterned by electron-beam lithography and ion etching from a bilayer constituted by a 20-nm thick film of Bi-doped Yttrium Iron Garnet (BiYIG) ($Bi_1Y_2Fe_5O_{12}$)[34] and a 6-nm thick Pt film. The waveguide is magnetized by a static magnetic field $H_0$. The field is applied perpendicular to the axis of the waveguide at an angle $\theta$ relative to its plane. The BiYIG film exhibits a strain-induced perpendicular magnetic anisotropy (PMA) with the effective anisotropy field $H_a$ = 1.8 kOe, which is close to the saturation magnetization of the film $4\pi M_s$ = 1.5 kG. This leads to a significant reduction of the ellipticity of magnetization dynamics and thus to a minimization of detrimental nonlinear spin-wave interactions[35]. The spin waves are inductively excited using a 300-nm wide and 80-nm thick Au antenna oriented perpendicular to the waveguide. The excitation microwave current with the carrier frequency $f$ is applied in the form of 100-ns wide pulses with the repetition period of 5 μs. Simultaneously, we apply 200-500-ns long pulses of dc current $I$ through the Pt layer on top of the magnetic waveguide. Due to the spin-Hall effect[36,37] in Pt, electrons with opposite orientations of the magnetic moment scatter toward opposite surfaces of the Pt film (see inset in Fig. 1a). As a result, the in-plane dc current in Pt is converted into an out-of-plane pure spin current $I_s$, which is injected into the BiYIG film. This injection results in a compensation of the natural magnetic damping in the magnetic material[22] and is expected to enable amplification of the propagating spin waves at sufficiently large $I$. The propagation of spin-wave pulses is visualized with high spatial and temporal resolution using the micro-focus Brillouin light scattering (BLS) spectroscopy[10] (see Methods for details). This technique yields a signal referred to as BLS intensity, which is proportional to the intensity of spin waves at the position where the probing laser light is focused. By synchronizing the detection of the scattered probing light with the excitation pulses, we obtain the possibility to directly analyze the propagation of spin-wave pulses in the space and time domain.

Figures 1c and 1d demonstrate the main result of our work – the true amplification of spin-wave pulses by the spin current, which is achieved at $H_0$ = 1.8 kOe applied at $\theta$ = 30°. As will be discussed below, the choice of these parameters is crucial to achieve the desired effect. Figure 1c shows the color-coded spin-wave intensity in the space-time coordinates for $I$=0 and 1.4 mA. The origin of the time axis is aligned with the beginning of the leading edge of the microwave pulse, while its span is equal to the pulse width of 100 ns. The origin of the space axis is aligned with the center of the antenna. The data obtained at $I$=0 demonstrate a damped propagation of the spin-wave pulse unaffected by the spin current. The leading edge of the pulse linearly shifts in time (dashed line), which corresponds to the constant group velocity of 135 m/s, while the intensity of the pulse strongly reduces between $x = 1$ and 10 µm due to the natural magnetic damping. In contrast, the data obtained at $I$=1.4 mA show a significant increase in the intensity of the pulse during propagation. We note that the group velocity remains unchanged in this regime. This indicates that the application of the dc current does not modify other parameters of the spin wave, except for damping, which obviously becomes negative.

Figure 1d illustrates the spatial attenuation/amplification of spin waves at different $I$. As seen from these data, at all currents, the spatial dependence of the spin-wave intensity is exponential (note the logarithmic scale of the vertical axis) and can be described by $e^{\kappa x}$, where $\kappa$ is the intensity decay constant. At $I = 0$, the intensity of the spin wave reduces by about two orders of magnitude at the propagation distance $x = 1$-10 µm. Taking into account the measured group velocity 135 m/s and the frequency of spin waves $f \approx 5.0$ GHz, this corresponds to the Gilbert damping parameter $\alpha \approx 1.2 \times 10^{-3}$, which is typical for BiYIG/Pt bilayers[38]. With the increase in $I$, the spatial attenuation becomes weaker. The attenuation completely vanishes at $I = 1.05$ mA, resulting in a decay-free propagation regime. Further increase in $I$ leads to an exponential increase of the intensity of spin waves in space. At the maximum current used in the experiment ($I = 1.4$ mA), the intensity increases by an almost a factor of 5 over the propagation distance $x = 1$-10 µm . Note that 1.4 mA corresponds to a current density of only $4 \times 10^{11}$ A/m$^2$. A significantly higher current could therefore be considered for future devices[39].

Although the results discussed above might lead to the conclusion that the amplification by spin current is straightforward to implement, a detailed consideration shows that this is not a trivial task and is possible only within a certain parameter window. Figure 2a shows the current dependence of the decay constant obtained for two orientations of the static magnetic field $\theta = 0$ and 30°. At $\theta = 30°$, the dependence is linear over the entire range $I = 0 - 1.4$ mA, as expected for the effects of the spin transfer torque on the magnetic damping[40]. It crosses zero at $I_C = 1.07$ mA, which marks the transition to the amplification regime. In contrast, at $\theta = 0$ the dependence is linear only at $I < I_C$. At larger currents, the decay constant saturates, and its value remains negative, indicating that the amplification regime is never achieved. Note that, while changing in the experiment $\theta$ from 0 to 30°, we simultaneously adjust the magnitude of $H_0$ from 2.0 to 1.8 kOe in order to maintain a constant frequency of the ferromagnetic resonance $f_{FMR} \approx 4.99$ GHz. This ensures that the experimental conditions are not strongly altered by the change in $\theta$.

We emphasize that previous studies of microscopic YIG/Pt waveguides without PMA also showed the absence of the amplification at currents exceeding the threshold for complete damping compensation[26,33]. As mentioned above, this was associated with the onset of intense magnetic auto-oscillations, leading to a strong scattering of propagating spin waves due to their nonlinear interaction with the auto-oscillations. Similarly, we also observe an onset of auto-oscillations in the BiYIG/Pt waveguide both at $\theta = 0$ and 30°, as shown in Figs. 2b and 2c, in which we observe an appearance of a narrow intense spectral peak at $I > 1$ mA indicating the excitation of auto-oscillations induced by the spin current. The main difference between the two cases lies in the nonlinear frequency shift of the auto-oscillation peak with increase of the auto-oscillation amplitude. While at $\theta = 0$, the auto-oscillation frequency noticeably increases with $I$, at $\theta = 30°$ it remains almost constant. These behaviors are caused by the interplay between the effects of PMA and the effects of the dipolar demagnetizing fields on the spectrum of magnetic excitations in the BiYIG film[38]. We emphasize that the vanishing nonlinear frequency shift is expected to be accompanied by a decrease in the ellipticity of the magnetization precession. This latter parameter is indeed known to be one of the most important factors for nonlinear coupling of spin waves responsible for their scattering[35].

In order to get better insight into the observed behaviors, we perform micromagnetic simulations using the MuMax3 software[41] (see Methods for details). First, we calculate the dispersion spectrum of spin waves in the waveguide for different angles $\beta$ of the magnetization precession cone and determine the nonlinear frequency shift. Figure 3a shows the nonlinear frequency shift calculated as the difference in the frequency of spin waves with zero wave vector obtained at $\beta = 10°$ and 0.1°, as a function of the angle of the static magnetic field $\theta$. These results are in good agreement with the experimental data (Figs. 2b and 2c). Both experiment and simulations show that the frequency shift is positive at $\theta = 0$ and almost vanishes with the increase of $\theta$ to 30°. Additionally, the simulations allow us to obtain information about the ellipticity of the magnetization precession, which is hardly accessible in the experiment. We calculate the ellipticity using the standard expression $\varepsilon = 1 - \frac{|m_{min}|^2}{|m_{max}|^2}$, where $m_{min}$ and $m_{max}$ are the smallest and the largest values of the dynamic magnetization over the precession cycle. Note that due to the complex distribution of the dynamic demagnetizing field across the width of the nano-patterned waveguide, the ellipticity differs significantly at the center and at the edges of the waveguide. As seen from Fig. 3b, it also shows essentially different angular dependences. At the center of the waveguide, the ellipticity is relatively small at $\theta = 0$ and slowly reduces with the increase in the angle. In contrast, at the edge of the waveguide, the ellipticity is as large as 0.5 at $\theta = 0$. This large ellipticity is generally expected to result in a strong nonlinear spin-wave coupling and is likely the reason for the nonlinear suppression of the effects of the spin current observed in the experiment at $\theta = 0$. As seen from Fig. 3b, the edge ellipticity quickly decreases with the increase in $\theta$, which can explain the absence of the adverse influence of intense auto-oscillations on the spin-current amplification observed in the experiment at $\theta = 30°$.

The data of simulations clearly show that shape effects in nano-patterned waveguides strongly influence the processes of amplification of propagating spin waves by pure spin currents by enhancing the nonlinear spin-wave scattering. In an extended magnetic film, the scattering can be efficiently suppressed by tuning the strength of PMA, which allows one to exactly compensate the dipolar demagnetizing fields and to obtain almost circular magnetization precession trajectory[35,42]. On the contrary, in the case of a patterned waveguide, one additionally needs to tune the angle of the static field to compensate for the shape effects. However, even this approach cannot provide a complete suppression of ellipticity over the entire cross section of the waveguide. As a result, nonlinear-shift management strategies alone cannot fully eliminate the detrimental nonlinear scattering of the propagating spin wave.

We develop an active strategy to achieve a full suppression of nonlinear scattering process and true spin-wave amplification based on a time synchronization between the spin-wave signal and the spin-current pulse. In particular, due to residual interaction of the signal spin wave with excited auto-oscillations, amplification can be achieved only within a certain time interval after the spin current is applied. Figure 4a shows representative temporal dependences of the intensity of magnetic oscillations after the start of the spin-current pulse. The dependence recorded at $I = 1.4$ mA corresponds to the case when the damping is completely compensated resulting in the development of auto-oscillations, while the dependence recorded at $I = 1.0$ mA corresponds to the case of incomplete compensation. Both dependences exhibit a rapid intensity increase (note the logarithmic scale of the vertical axis) directly after the start of the dc pulse. This increase reflects the known effect of the enhancement of magnetic fluctuations by the spin current[43] and does not indicate an onset of auto-oscillations. At $I = 1.4$ mA, the initial increase is followed by an exponential growth of the intensity (dashed line in Fig. 4a). The growth rate linearly depends on $I$ (Fig. 4b), as expected for auto-oscillations induced by the spin-transfer torque[40]. Note that the linear dependence in Fig. 4b crosses zero at $I=1.07$ mA, which is consistent with the critical current necessary to observe the amplification of propagating spin waves (Fig. 2a).

Similarly to any system exhibiting auto-oscillations, the exponential growth of auto-oscillations is followed by saturation caused by an increase in the nonlinear scattering (nonlinear damping) at large oscillation amplitudes[40]. To avoid this nonlinear scattering process, the spin-wave amplification should happen before the saturation takes place, as demonstrated by the data in Fig. 4c. The figure shows the spatial dependences of the intensity of the signal spin wave obtained from measurements performed with spin-wave pulses propagating either during the transient-regime interval or in the saturation steady-state regime. As seen from these data, we observe the true spin-wave amplification only in the transient regime and an attenuation in the saturation regime. This key result evidences that, even in systems with minimized nonlinear scattering, efficient amplification is possible only with spin-current pulses that are short enough to not drive the spin system into the strongly nonlinear saturated state.

Finally, we discuss the robustness of the amplification regime by changing the frequency and the group velocity of spin waves. We vary the frequency of the microwave signal applied to the antenna and determine the group velocity $v_g$ of the spin-wave pulses (Fig. 5a). As seen from these data, the group velocity increases by about a factor of two in the frequency range 5.0-5.2 GHz. In Fig. 5b, we then show the frequency dependences of the spatial spin-wave decay constant obtained in the attenuation regime (point-down triangles) and in the amplification regime (point-up triangles). The twofold decrease in the decay constant observed in the attenuation regime, is a natural consequence of the twofold increase in the group velocity. A less obvious result is the observed decrease in the magnitude of the decay constant in the amplification regime, indicating less efficient amplification of faster waves. However, this result can be reproduced using a simple theoretical model. We calculate the spatial decay constant as

$$\kappa = \Gamma_r / 2v_g, \tag{1}$$

where $\Gamma_r$ is the temporal relaxation rate of the magnetization precession, which is determined from the data of Fig. 4b. The obtained dependencies (solid curves in Fig. 5b) agree well with the experimental data.

In conclusion, our results provide direct experimental evidence for the possibility of the true spatial amplification of propagating spin waves by spin currents in magnonic nano-waveguides. They unambiguously identify the physical phenomena that prevented amplification in previous experimental studies and show how these detrimental effects can be overcome in practice. These findings open new avenues for the field of nano-magnonics by demonstrating a simple and energy-efficient approach for the on-chip amplification of propagating spin waves, which can be used in most of nanoscale magnonic devices. The possibility to directly amplify propagating spin waves with a small dc current enables the implementation of complex cascadable magnonic nano-circuits with a large fan-out that do not require the energy-consuming conversion of spin waves into radio-frequency electronic signals for compensation of propagation losses, which is expected to significantly advance the practical realization of magnon-based computing platforms. Furthermore, clocked amplification using short current pluses of low intensity is fully compatible with CMOS operation enabling the development of cost-effective CMOS/magnon chips.

**Data availability**

The datasets generated during and/or analyzed during the current study are available from the corresponding author on reasonable request.

**Methods**

**Sample fabrication.**

The BiYIG film was grown by pulsed laser deposition (PLD) on substituted Gallium Gadolinium Garnet (sGGG) substrate using stoichiometric BiYIG target. The distance between the target and the substrate was 44 mm. The deposition was performed using a frequency tripled Nd:YAG laser ($\lambda = 355$ nm) with a 2.5-Hz repetition rate and a fluency of about 1 J/cm$^2$. The uniaxial anisotropy is set by the choice of the substrate temperature[42] (420 °C). Prior to the deposition, the substrate was annealed at 700°C under 0.4 mbar of $O_2$. The growth was performed at 0.25 mbar $O_2$ pressure. Finally, the sample was cooled down under 300 mbar of $O_2$. No post annealing was performed. The Pt layer was deposited using dc magnetron sputtering. Prior to Pt deposition, the BiYIG film was slightly etched with $O_2$ in order to remove photo-resist residues and promote surface spin-transparency.

**Micro-focus BLS measurements.** The measurements were performed at room temperature. The detection of propagating spin waves and current-induced auto-oscillations was performed using the analysis of the inelastic scattering of laser light from magnetic excitations. The probing laser light had the wavelength of 532 nm and the power of 0.1 mW. The light was focused through the transparent sGGG substrate onto the BiYIG film using a high-performance corrected microscope objective lens with the magnification of 100 and the numerical aperture of 0.85. The light scattered from magnetic excitations was collected by the same lens and sent for analysis to a six-pass Fabry-Perot interferometer. The intensity of the scattered light (BLS intensity) was proportional to the intensity of magnetization oscillations at the position of the probing spot. The temporal resolution was achieved by synchronizing the detection of the scattered light with the excitation of spin waves. By moving the probing laser spot along the waveguide, detection of the spin-wave intensity with simultaneous spatial and temporal resolution was implemented. To achieve high spatial accuracy (<50 nm), active stabilization of the sample position was used.

**Micromagnetic simulations.** We numerically simulated spin-wave dynamics in a 500 nm wide and 20-nm thick strip waveguide with the length $L$=20 μm. The computation domain was discretized into 10 nm × 10 nm × 10 nm cells with periodic boundary conditions at the ends of the waveguide. The magnetization dynamics was excited by spatially-periodic deflection of magnetic moments from their equilibrium orientation in the direction parallel to the waveguide axis. The spatial period of the deflection defined the wavelength of the excited spin waves, and the deflection angle defined the angle (amplitude) of the magnetization precession. We analysed the free dynamics of the magnetization caused by the initial deflection and determined the frequency corresponding to a given wavelength and the amplitude of spin waves. This approach allowed us to calculate the amplitude-dependent dispersion spectrum of spin waves and its nonlinear shift caused by an increase in the precession angle.

**Corresponding author:** Correspondence and requests for materials should be addressed to V.E.D. ([demidov@uni-muenster.de](demidov@uni-muenster.de)).

**Acknowledgements**

This work was supported by in part by the Deutsche Forschungsgemeinschaft (DFG, German Research Foundation) – project number 423113162, by the ANR MAESTRO project, Grant No. 18-CE24-0021 of the French Agence Nationale de la Recherche, Labex NanoSaclay "SPICY" ANR-10-LABX-0035, and received financial support from the Horizon 2020 Framework Program of the European Commission under FET-Open grant agreement no. 899646 (k-NET). H. M. thanks L. Thevenard for useful discussions.

**Author Contributions:** H.M. performed the nanofabrication, measurements, and data analysis. B.D. performed measurements and data analysis. D.G. grew and characterized the films. A.E.-K. performed data analysis. R.L., V.C., and P.B. contributed to the design and implementation of the research. A.A., S.O.D. and V.E.D. formulated the experimental approach and supervised the project. All authors co-wrote the manuscript.

**Competing interests**

The authors have no competing financial interests.

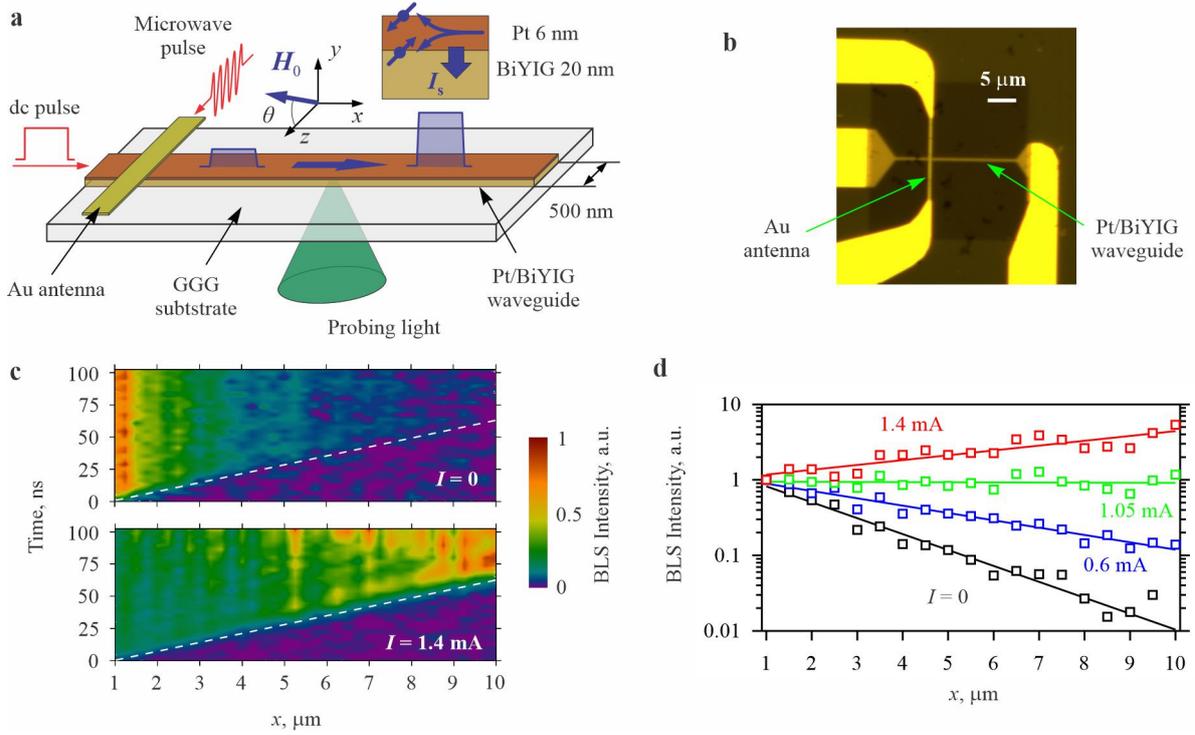

**Figure 1 Implementation of spin-wave amplification** (**a**) Schematics of the experiment. Spin-wave pulses are excited by a Au antenna and propagate in a 500-nm wide waveguide fabricated from a BiYIG(20 nm)/Pt(6 nm) bilayer. In-plane dc current flowing in the Pt layer is converted into an out-of-plane pure spin current $I_s$ (inset), which is injected into the BiYIG film and exerts anti-damping torque on the magnetization. (**b**) Optical micrograph of the sample. (**c**) Normalized BLS maps of the spin-wave intensity in the space-time coordinates recorded at $I = 0$ and 1.4 mA, as labeled. Dashed lines show the spatio-temporal shift of the edge of the spin-wave pulse corresponding to the group velocity of 135 m/s. (**d**) Spatial dependence of the intensity of the spin-wave pulse measured at different dc currents, as labeled. Symbols show the experimental data. Solid straight lines show the exponential fit. The data are obtained at $f = 5.025$ GHz and $H_0 = 1.8$ kOe applied at $\theta = 30°$.

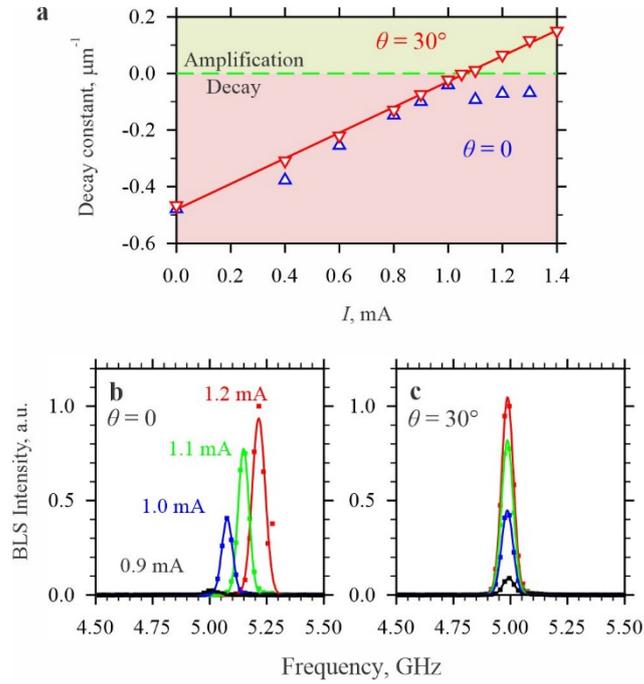

**Figure 2 | Effect of the angle of the static magnetic field on amplification and auto-oscillations.** (**a**) Current dependence of the decay constant of spin-wave intensity obtained at $\theta = 0$ and $30°$, as labeled. Symbols show experimental data. Solid line is the linear fit of the data at $\theta = 30°$. (**b**) and (**c**) BLS spectra of magnetization auto-oscillations recorded at the labeled values of the dc current at $\theta = 0$ and $30°$, respectively. The data for $\theta = 0$ and $30°$ were obtained at $H_0 = 2.0$ and $1.8$ kOe, respectively, to compensate for the frequency shift of the spin-wave dispersion spectrum.

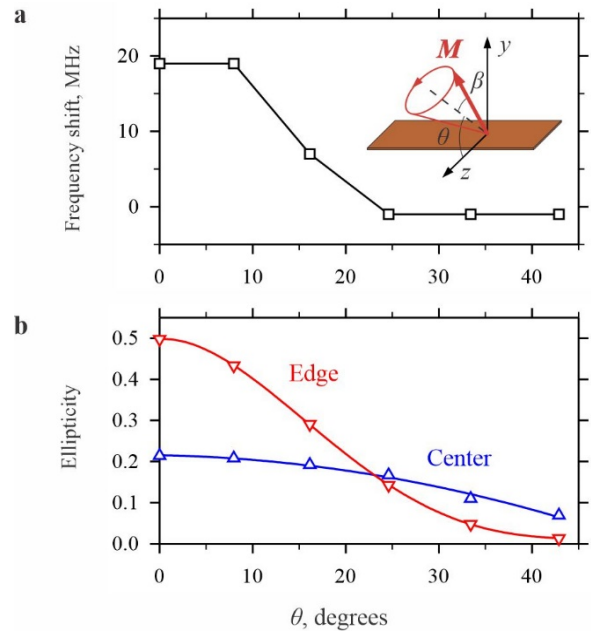

**Figure 3 | Results of micromagnetic simulations of spin-wave dynamics in the nano-waveguide.** (**a**) Angular dependence of the nonlinear frequency shift of the spin-wave spectrum calculated as the difference in the frequency of spin waves for the magnetization precession cone angles $\beta$ of 10° and 0.1°. (**b**) Angular dependences of the ellipticity of the magnetization precession in the center and at the edge of the nano-waveguide. Curves are guides for the eye. These simulations points toward $\theta = 30°$ as a field angle, at which both the nonlinear frequency shift and the ellipticity almost vanish.

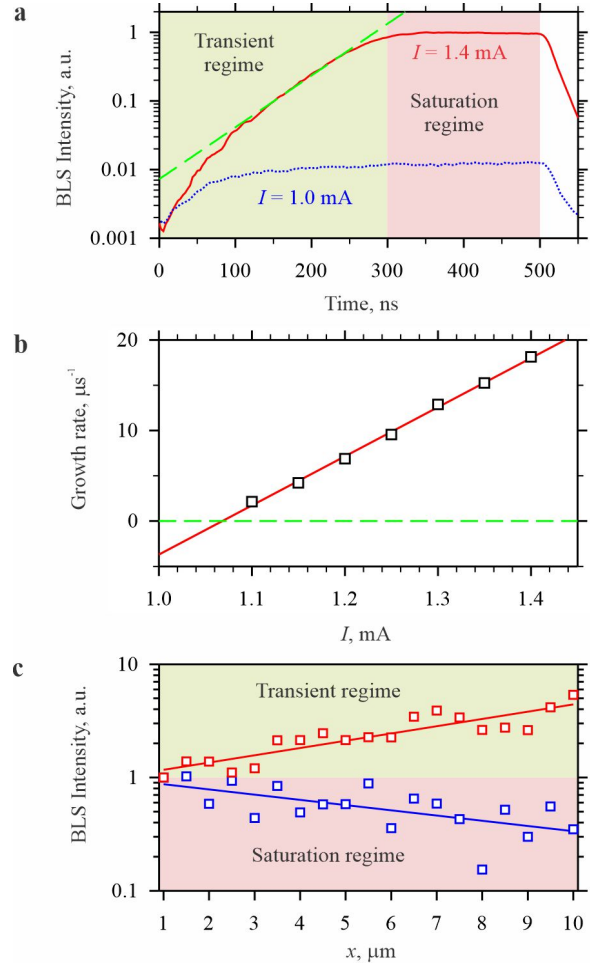

**Figure 4 | Time constraints of the amplification process.** (**a**) Temporal dependences of the intensity of auto-oscillations after the start of the dc pulse recorded at $I = 1.4$ mA (solid curve). The dependence shows the transient regime and the saturation regime with a cross over at about 300 ns. Dashed curve shows the exponential increase of the intensity of auto-oscillations. Temporal dependence of the intensity of magnetic fluctuations (dotted curve) recorded at $I = 1.0$ mA is shown for reference. (**b**) Current dependence of the exponential growth rate of current-induced auto-oscillations. Symbols show the experimental data. Line is a linear fit. (**c**) Spatial dependences of the intensity of the signal spin wave for the cases when the spin-wave pulse is applied during the time interval corresponding to the transient and saturation regime, as labeled. Symbols show the experimental data. Solid lines show the exponential fit. The data are obtained at $f = 5.025$ GHz and $H_0 = 1.8$ kOe applied at $\theta = 30°$.

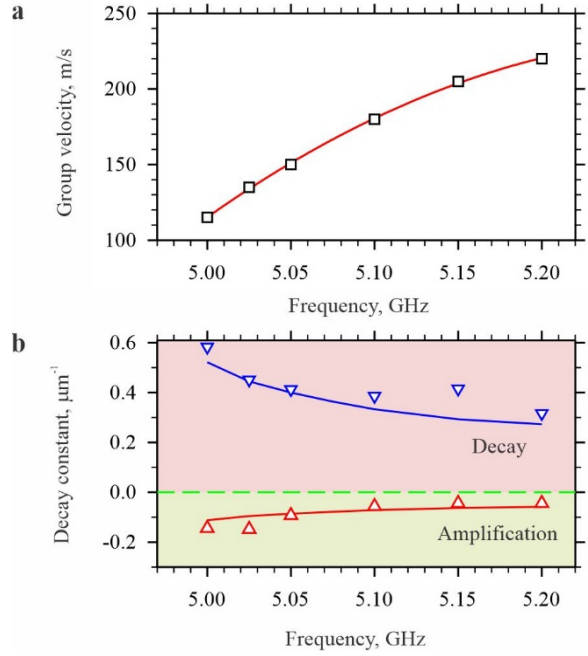

**Figure 5 | Effects of the velocity of spin waves on the amplification efficiency.** (**a**) Frequency dependence of the group velocity. Symbols show the experimental data. Curve is the guide for the eye. (**b**) Frequency dependences of the decay constant obtained at $I = 0$ (point-down triangles) and at the maximum current (point-up triangles). Note that negative decay constants correspond to the spatial amplification of the wave. Curves show the results of calculations using Eq. (1). The data are obtained at $H_0 = 1.8$ kOe applied at $\theta = 30°$. The largest amplification efficiency is obtained for the slowest spin-waves